\documentclass[twocolumn,showpacs,pra,aps,amssymb]{revtex4}
\usepackage{graphicx}
\begin{document}
\title{\bf General entanglement}

\author{Alexander A. Klyachko and Alexander S. Shumovsky}

\affiliation{Faculty of Science, Bilkent University, Bilkent,
Ankara, 06800 Turkey}

\begin{abstract}
The paper contains a brief review of an approach to quantum
entanglement based on analysis of dynamic symmetry of systems and
quantum uncertainties, accompanying the measurement of mean value
of certain basic observables. The latter are defined in terms of
the orthogonal basis of Lie algebra, corresponding to the dynamic
symmetry group. We discuss the relativity of entanglement with
respect to the choice of basic observables and a way of
stabilization of robust entanglement in physical systems.
\end{abstract}

\pacs{03.65.Ud, 03.67.-a}

\maketitle

\section{Introduction}

Entanglement, which is considered nowadays as the main physical
resource of quantum information processing and quantum computing,
has been discovered as a physical phenomenon representing ``the
characteristic trait of quantum mechanics" (Schr\"{o}dinger 1935).

According to the modern point of view, entangled states form a
special class of quantum states closed under SLOCC (Stochastic
Local Operations assisted by Classical Communications) (D\"{u}r
{\it et al} 2000, Verstraete {\it et al} 2002, Miyake 2003). Two
states belong to the same class iff they are converted into each
other by SLOCC. Mathematically SLOCC amounts to action of the
complexified dynamic symmetry group $G^c$ of the system
(Verstraete {\it et al} 2002). This
%classification can be
%naturally associated with the
description puts entanglement in general framework of geometric
invariant theory and allows extend it to arbitrary quantum systems
(Klyachko 2002).

SLOCC cannot transform entangled state into unentangled one and
vice versa (D\"{u}r {\it et al} 2000).  We define {\it completely
entangled\,} (CE) states, manifesting maximal entanglement in
their SLOCC class,
%Thus, for a given system, the class of entangled states can
%be specified by a certain {\it generic state}, manifesting {\it
%complete entanglement} (CE),
such that all entangled states of a given system can be
constructed from them by means of SLOCC.

It was shown recently that CE states manifest the maximal amount
of quantum fluctuations
%peculiar to the measurement of mean values of {\it basic} physical observables
(Can {\it et al} 2002(a), Klyachko and Shumovsky 2003, Klyachko
and Shumovsky 2004). This property can be used as a physical
definition of CE states.

It should be stressed that
%the existence of
quantum fluctuations caused by the representation of observables
in terms of Hermitian operators is an undoubted ``characteristic
trait" of quantum systems. Within the classical description, the
observables should be associated with c-numbers and hence are
incapable of manifestation of quantum fluctuations.

We now note that characterization of quantum states with respect
to quantum fluctuation is a common way in quantum optics. Coherent
(Glauber 1963, Perelomov 1986) and squeezed (Stoler 1970, Dodonov
2002) states provide an important examples. In particular, it has
been recognized recently that coherent states can in general be
associated with the unentangled (separable) states (Klyachko 2002,
Barnum {\it et al} 2003). In turn, there are also attempts to
characterize entanglement in terms of quantum fluctuations.

The aim of this article is to discuss the corollaries coming from
the physical definition of CE states via quantum fluctuations. We
mostly concentrate on the {\it relativity of entanglement} and on
the creation of {\it robust entanglement}. Let us emphasize once
more that as soon as CE states are defined, all other entangled
states of the same system can be obtained from CE states by means
of SLOCC.

The paper is arranged as follows. In Sec. 2 we briefly discuss the
specification of basic observables based on the consideration of
the dynamic symmetry properties of quantum systems and express the
definition of CE states in terms of a variational principle. Then,
in Sec. 3 the relativity of entanglement with respect to the
choice of basic observables is considered. In Sec. 4 we discuss
the stabilization of entanglement. Finally, in Sec. 5 we briefly
summarize the obtained results.

\section{Basic observables}

Quantum entanglement as well as any other quantum phenomenon
manifests itself via measurement of physical observables (Bell
1966).
%According to von Neumann's approach to quantum
%measurements, an observable property of a quantum system is given
%by a selfadjoint bounded linear operator on a certain Hilbert
%space (von Neumann 1996).
In von Neumann approach (von Neumann 1996) all observables
%$X_A:\mathcal{H}_A\rightarrow\mathcal{H}_A$ or what is the same
%all unitary transformations
%$e^{itX_A}:\mathcal{H}_A\rightarrow\mathcal{H}_A$
are supposed to be equally accessible. However physical nature of
the system often imposes inevitable constraints.
%usually
%imposes some restrictions on possible manipulation with quantum
%states.

For example, the components of composite system
$\mathcal{H}_{AB}=\mathcal{H}_A\otimes\mathcal{H}_B$ may be
spatially   separated by tens of kilometers,  as in EPR pairs used
in quantum cryptography. In such circumstances only local
observations $X_A$ and $X_B$ are available.

As another example, consider a system of $N$ identical particles,
each with space of internal degrees of freedom $\mathcal{H}$. By
Pauli principle
%imposes strong constraints onto accessible quantum states the
the state space of such system shrinks to {\it symmetric
tensors\,} $S^N\mathcal{H}\subset\mathcal{H}^{\otimes N}$ for
bosons, and to {\it skew symmetric tensors\,}
$\wedge^N\mathcal{H}\subset\mathcal{H}^{\otimes N}$  for fermions.
This superselection rule imposes severe restricion on manipulation
with quantum states, effectively reducing
% Because of
%indistinguishability of the particles
the accessible measurements
%observables
%upon
%for such system are the same as for
to that of a single particle.

This consideration led many researchers to the conclusion, that
available observables  should be included in description of any
quantum system from the outset, see Hermann 1966, Emch 1984.
Robert Hermann 1966 stated this thesis as follows
\begin{quote}
{\it ``The basic principles of quantum mechanics seem to require
the postulation of a Lie algebra of observables and a
representation of this algebra by skew-Hermitian operators."}
% R. Hermann, Lie Groups for physicists, New York, Benjamin, 1966
\end{quote}
We denote this  {\it Lie algebra of observables\,} by $\frak{L}$.
The corresponding Lie group
$$
G=\exp (i\frak{L})
$$
will be called {\it dynamic symmetry group\,} of the system. We'll
refer
%to state space  $\mathcal{H}$ together with
to unitary representation of the dynamical group $G$ in state
space $\mathcal{H}_S$ as {\it quantum dynamical system}.

%(e.g., see Emch 1984 and Brading and Castellani 2003; the latter
%reference represents a collection of fundamental papers on
%symmetries in physics).

Note finally that there is no place for entanglement in von
Neumann picture, where full dynamical group
$\mathrm{SU}(\mathcal{H})$
%ensures that
makes all states equivalent. Entanglement is an effect caused by
superselection rules or symmetry breaking which reduce the
dynamical group to a subgroup $G\subset \mathrm{SU}(\mathcal{H})$
small enough to create intrinsic   difference between states. For
example, entanglement in two component system
$\mathcal{H}_A\otimes\mathcal{H}_B$ comes from reduction of the
dynamical group to
$\mathrm{SU}(\mathcal{H}_A)\times\mathrm{SU}(\mathcal{H}_B)\subset
\mathrm{SU}(\mathcal{H}_A\otimes\mathcal{H}_B)$. Entanglement
essentially depends on the dynamical group and {\it must\,} be
discussed in framework of a given quantum dynamical system
$G:\mathcal{H}$. This {\it relativity of entanglement} is one of
the topics of this paper.

%The physical definition of CE states in terms of quantum
%fluctuations (Klyachko and Shumovsky 2004) requires specification
%of {\it basic observables} for a given quantum system.
For calculations we choose an arbitrary orthonormal basis $X_i,
i=1\ldots N$ of $\frak{L}=\mathrm{Lie}(G)$ and call its elements
$X_i$ {\it basic observables\,} (Klyachko 2002 and Klyachko and
Shumovsky 2003).

For example, in the case of a qubit (spin-$\frac{1}{2}$
``objects") the dynamic symmetry group is $G=\mathrm{SU}(2)$ and
$\frak{L}=\frak{su}(2)$ is algebra of traceless Hermitian $2\times
2$ matrices. One can choose spin projector operators $J_x,J_y,J_z$
(or the Pauli matrices) as the basic observables.
%Any observable
%(a $2
%\times 2$ Hermitian matrix) can be represented as a linear
%combination of these operators and an irrelevant scalar operator.

%The quantum fluctuations are described by the uncertainties of
%measurement of the mean values of observables.
%If $X_i$ is a basic
%observable and $\psi \in \mathcal{H}_S$ is a state of a system $S$
%determined in the Hilbert space $\mathcal{H}_S$,  the
%corresponding uncertainty of a measurement of the mean value
%$\langle \psi |X_i|\psi \rangle$ in the state $\psi$ is given by
%the variance

The level of quantum fluctuations of a basic observable $X_i$ in
state $\psi\in\mathcal{H}_S$ of system $S$ is given by the
variance
\begin{eqnarray}
\mathbb{V}(X_i,\psi)= \langle \psi|X_i^2|\psi \rangle - \langle
\psi|X_i|\psi \rangle^2 \geq 0. \label{1}
\end{eqnarray}
Summation over all basic observables of the quantum dynamic system
gives the {\it total uncertainty} (total variance) peculiar to the
state $\psi$:
\begin{eqnarray}
\mathbb{V}(\psi)= \sum_{i} \mathbb{V}(X_i, \psi) = \sum_{i}\langle
\psi|X_i^2|\psi \rangle - \langle \psi|X_i|\psi \rangle^2 .
\label{2}
\end{eqnarray}
This quantity is independent of the choice the basic observables
and measures the total level of quantum fluctuations in the
system.

%An important physical case is provided by the basic observables
%corresponding to a compact Lie algebra. Then, there is
Recall that the Casimir operator
\begin{eqnarray}
\widehat{C}= \sum_i X_i^2 , \nonumber
\end{eqnarray}
which appears in Eq. (2) is independent of  the choice of the
basis $X_i$ and acts as a multiplication by scalar $C$ if
representation $G:\mathcal{H}_S$ is irreducible.
%%where
%$\mathbf{1}$ denotes the unit operator and $C$ is a positive
%number.
In this case Eq. (2) takes the form
\begin{eqnarray}
\mathbb{V}(\psi) = C- \sum_i \langle \psi |X_i| \psi \rangle^2 .
\label{3}
\end{eqnarray}

It has been observed that completely entangled (CE) states of an
arbitrary number $n \geq 2$ of qubits obey a certain conditions.
Namely, expectation values of all three spin-projection operators
for all parties of the system have zero value in CE state
$|\psi_{CE} \rangle$ (Can {\it et al} 2002). In general, the
condition
\begin{eqnarray}
\forall i \quad \langle \psi_{CE}|X_i|\psi_{CE} \rangle =0, \quad
|\psi_{CE} \rangle \in \mathcal{H}_S, \label{4}
\end{eqnarray}
can be used as a general physical definition of CE (Klyachko and
Shumovsky 2004). This is an {\it operational} definition of CE
(definition in terms of what can be directly measured).

%Since the form (1) is nonnegative by definition for any $|\psi\rangle \in \mathcal{H}_S$,
From Eq. (3) it follows that the total variance  attains  its
maximal value  equal to Casimir in the case of CE states:
\begin{eqnarray}
\mathbb{V}(\psi_{CE})= \max_{\psi \in \mathcal{H}_S}
\mathbb{V}(\psi)=C. \label{5}
\end{eqnarray}
This Eq. (5) is, in a sense, equivalent to the maximum of entropy
principle, defining the equilibrium states in quantum statistical
mechanics (Landau and Lifshitz 1980).

From operational point of view state $\psi\in\mathcal{H}$ is
entangled if one can prepare a completely entangled state
$\psi_{CE}$ from it using SLOCC operations. It should be
emphasized that SLOCC transformations have been identified with
action of the {\it complexified dynamical group}
$$G^c=\exp(\frak{L}\otimes\mathbb{C}),$$
%are associated with the operations from the complexified dynamic
%symmetry group
of the system (Verstraete 2003).
%If now $g^c \in G^c$ denotes SLOCC, then the
This leads us to the following definition of general entangled
states $\psi_E$ of the system
\begin{eqnarray}
\psi_{E}  = g^c\psi_{CE}  ,\quad \mbox{for some }g^c\in G^c.
\label{6} \end{eqnarray}
 Thus, {\it the general entangled states
can be defined as that obtained from the states, manifesting
maximum total uncertainty, by action of the complexified dynamic
group.}

Let's stress that the above definition of basic observables and
equations of CE (4) do not assume the composite nature of the
system $S$. In other words, a single-particle system can manifest
entanglement if its state obeys the conditions (4) (Can {\it et
al} 2005).

\section{Relativity of entanglement}

%If in the system $S$ the number of independent states per party
%exceeds two, there can be more than one nontrivial dynamic
%symmetry group.

Physics of quantum system $S$ with given Hilbert state space
$\mathcal{H}_S$ may implies different dynamical groups.

An important example is provided by a {\it qutrit} (three-state
quantum system), which is widely discussed in the context of
quantum ternary logic (Bechman-Pasquinucci and Peres  2000,
Bru{\ss} and Macchiavello 2002, Kaszlikowski {\it et al} 2003). In
this case, the general symmetry is given by $G=\mathrm{SU}(3)$, so
that the local basic observables are given by the eight
independent Hermitian generators of the $\frak{L}=\frak{su}(3)$
algebra (see Caves and Milburn 2000). In the special case of
spin-1 system, the symmetry is reduced to the $G'=\mathrm{SU}(2)$
group, and the corresponding local basic observables coincide with
the three spin-1 operators (Can {\it et al} 2005). Since
$\frak{su}(2) \subset \frak{su}(3)$, the qutrit entanglement with
respect to $\frak{su}(3)$ observables implies entanglement in the
$\frak{su}(2)$ domain but not vice versa.

For example, a single spin-1 object can be entangled with respect
to the $\frak{su}(2)$ basic observables  but not in the
$\frak{su}(3)$ sector (Can {\it et al} 2005). A general spin-1
state has the form
\begin{eqnarray}
|\psi \rangle = \sum_{s=-1}^1 \psi_s |s \rangle , \quad \sum_s
|\psi_s|^2 =1, \label{7}
\end{eqnarray}
where $s=0, \pm 1$ denotes the spin projection. In the basis $|s
\rangle$, the spin-1 operators have the form
\begin{eqnarray}
S_x= \frac{1}{\sqrt{2}} \left( \begin{array}{ccc} 0 & 1 & 0 \\ 1 &
0 & 1 \\ 0 & 1 & 0 \end{array} \right) , \quad S_y=
\frac{i}{\sqrt{2}} \left( \begin{array}{crr} 0 & -1 & 0 \\ 1 & 0 &
-1 \\ 0 & 1 & 0 \end{array} \right) \nonumber \\ S_z= \left(
\begin{array}{ccr} 1 & 0 & 0 \\ 0 & 0 & 0 \\ 0 & 0 & -1 \end{array}
\right). \nonumber
\end{eqnarray}
Using CE condition (4) with the basic observables $S_i$ and taking
into account the normalization condition in (7), we obtain four
equations for six real parameters $\mathrm{Re}( \psi_s)$ and
$\mathrm{Im}( \psi_s)$. In particular, the state with zero
projection of spin $|0 \rangle$ manifests CE. This state $|0
\rangle$ together with the states
\begin{eqnarray}
\frac{1}{\sqrt{2}} (|1 \rangle \pm |-1 \rangle ), \nonumber
\end{eqnarray}
form the basis of CE states in the three-dimensional Hilbert space
of spin-1 states. The possibility of the single spin-1
entanglement was also discussed by Viola {\it et al} (Viola {\it
et al} 2004).

To understand the physical meaning of this CE, we note that there
is a certain correspondence between the states of two qubits and
single qutrit provided by the Clebsch-Gordon decomposition
\begin{eqnarray}
\mathcal{H}_{\frac{1}{2}} \otimes
\mathcal{H}_{\frac{1}{2}}=\mathcal{H}_1 \oplus \mathcal{H}_0 .
\nonumber
\end{eqnarray}
Here $\mathcal{H}_{\frac{1}{2}}$ denotes the two-dimensional
Hilbert space of a single qubit. The three-dimensional Hilbert
space $\mathcal{H}_1$ contains the symmetric states of two qubits
\begin{eqnarray}
|1 \rangle  = |\uparrow \uparrow \rangle , \quad |0 \rangle =
\frac{1}{\sqrt{2}} (|\uparrow \downarrow \rangle +|\downarrow
\uparrow \rangle ), \quad |-1 \rangle =|\downarrow \downarrow
\rangle \label{8}
\end{eqnarray}
while $\mathcal{H}_0$ corresponds to the antisymmetric state
\begin{eqnarray}
|A \rangle= \frac{1}{\sqrt{2}} (|\uparrow \downarrow \rangle
-|\downarrow \uparrow \rangle ). \label{9}
\end{eqnarray}
It is now seen that the state $|0 \rangle$ of spin-1 is CE in
terms of a certain pare of spin-$\frac{1}{2}$ ``particles", which
can be interpreted as intrinsic degrees of freedom for the spin-1
object.

A vivid physical example is provided by the $\pi$-mesons. It is
known that three $\pi$-mesons form an isotriplet (Bogolubov and
Shirkov 1982)
\begin{eqnarray}
\pi^+=|1 \rangle, \quad \pi^0=|0 \rangle , \quad \pi^-=|-1
\rangle, \label{10}
\end{eqnarray}
where $|\ell \rangle$ ($\ell =0, \pm 1$) denotes the states of
isospin $I=1$. From the symmetry point of view, isospin is also
specified by the $\mathrm{SU}(2)$ group. Thus, in view of our
discussion one can conclude that $\pi^0$ meson is CE with respect
to internal degrees of freedom.

The internal structure of mesons is provided by the quark model
(Huang 1982). Namely, the fundamental representation of the
isospin symmetry corresponds to the two doublets (qubits) that
contain the so-called up ($u$) and down ($d$) quarks and
anti-quarks ($\bar{u}$ and $\bar{d}$). In terms of quarks, the
isotriplet (10) has the form
\begin{eqnarray}
\pi^+ =u \bar{d},  \quad \pi^0= \frac{1}{\sqrt{2}} (u \bar{u}+d
\bar{d}), \quad \pi^-= \bar{u} d. \nonumber
\end{eqnarray}
It is now clearly seen that $\pi^0$ meson represents CE state with
respect to quark degrees of freedom. An oblique corroboration of
this fact is given by the high instability of $\pi^0$ meson in
comparison with $\pi^{\pm}$. Such an instability may result from
the much higher amount of quantum fluctuations peculiar to CE
state.

Another example is given by a single dipole photon, which is
emitted by a dipole transition in atom or molecule and carries
total angular momentum $J=1$ (Berestetskii {\it et al} 1982). In
the state with projection of the total angular momentum $m=0$ it
is completely entangled. In fact, such photon carries two qubits.
One of them is the polarization qubit, which is usually considered
in the context of quantum information processing. Another qubit is
provided by the orbital angular momentum, which can be observed
(Padgett {\it et al} 2002) and used for the quantum information
purposes (Mair {\it et al} 2001). Like in the case of $\pi^0$
meson, these two qubits correspond to the intrinsic degrees of
freedom of the photon.

As one more example, let us consider the so-called {\it biphoton},
which consists of two photons of the same frequency, created at
once, and propagating in the same direction (Burlakov {\it et al}
1999, Chechova {\it et al} 2004). Before splitting, biphoton can
be interpreted as a single ``particle". In the basis of linear
polarizations, the states of biphoton have the form
\begin{eqnarray}
\left\{ \begin{array}{lcl} |1 \rangle  & = &  |x,x \rangle  \\
|0 \rangle  & = & \frac{1}{\sqrt{2}} (|x,y \rangle +|y,x \rangle )
\\ |-1 \rangle & = & |y,y \rangle \end{array} \right. \label{11}
\end{eqnarray}
(the propagation direction is chosen as the $z$-axis). Thus,
formally they coincide with the spin-1 states. It should be
stressed that the antisymmetric state
\begin{eqnarray}
|A \rangle = \frac{1}{\sqrt{2}} (|x,y \rangle - |y,x \rangle ),
\nonumber
\end{eqnarray}
is forbidden (Berestetskii {\it et al} 1982). The CE of the state
$|0 \rangle$ in (11) is evident.

The antisymmetric state is also forbidden in a system of two
two-level atoms with dipole interaction in the Lamb-Dicke limit of
short distances (\c{C}ak{\i}r {\it et al} 2005), so that this
system can also be considered as a single spin-1 object.

Although a single qutrit can be prepared in CE state in the
$\mathrm{SU}(2)$ sector, it does not manifest entanglement in the
$\mathrm{SU}(3)$ sector, where the local observables are given by
the eight independent Hermitian generators of the $\frak{su}(3)$
algebra (Caves and Milburn 2000):
\begin{eqnarray}
 \left( \begin{array}{ccc} 0 & 1 & 0 \\ 1 & 0 &  0 \\ 0 & 0 & 0
\end{array} \right), & \left( \begin{array}{crr} 0 & -i & 0 \\ i &
0 & 0 \\ 0 & 0 & 0 \end{array} \right) , & \left(
\begin{array}{crr} 1 & 0 & 0 \\ 0 & -1 & 0 \\ 0 & 0 & 0
\end{array} \right) \nonumber \\ \left( \begin{array}{ccc} 0 & 0 & 1 \\ 0 & 0 & 0 \\ 1 &
0 & 0 \end{array} \right) , & \left( \begin{array}{ccr} 0 & 0 & -i
\\ 0 & 0 & 0 \\ i & 0 & 0 \end{array} \right) , & \label{12} \\
\left( \begin{array}{ccc} 0 & 0 & 0 \\ 0 & 0 & 1 \\ 0 & 1 & 0
\end{array} \right) , & \left( \begin{array}{ccr} 0 & 0 & 0 \\ 0 &
0 & -i \\ 0 & i & 0 \end{array} \right) , & \frac{1}{\sqrt{3}}
\left( \begin{array}{ccr} 1 & 0 & 0 \\ 0 & 1 & 0 \\ 0 & 0 & -2
\end{array} \right) . \nonumber
\end{eqnarray}
It is easily seen that conditions (6) cannot be realized for the
state (7) with the basic observables (12).

Thus, a single three-state quantum system (qutrit) may or may not
manifest entanglement, depending on what kind of basic observables
is accessible. Hence, there is a relativity of entanglement with
respect to choice of basic observables.

\section{Stabilization of entanglement}

Numerous applications of quantum entanglement require not an
arbitrary entangled state but a {\it robust} one. This assumes the
high amount of entanglement together with the long lifetime of the
entangled state. This lifetime is usually determined by
interaction with a dissipative environment, which causes the
decoherence in the system.

The approach under discussion reveals a way of obtaining robust
entanglement. In conformity with the definition (5), we should
first prepare a state of a given system with the maximal amount of
quantum fluctuations of all basic observables. As the second step,
we should decrease the energy of the system up to a minimum (local
minimum) to stabilize the state, keeping the level of quantum
fluctuations. Thus obtained state would be stabile (metastable)
and CE.

As an example, consider atomic entanglement caused by photon
exchange between two atoms in a cavity. In the simplest case of
two-level atoms in an ideal cavity, containing a single photon,
the CE state in atomic subsystem arises and decays periodically
due to the Rabi oscillations (Plenio {\it et al} 1999). The above
stabilization scheme can be used if instead we consider
three-level atoms with the $\Lambda$-type transitions (Can {\it et
al} 2002(b), Can {\it et al} 2003).

In this case, the two three-level atoms with allowed dipole
transitions $1 \leftrightarrow 2$ and $2 \leftrightarrow 3$ and
dipole forbidden transition $1 \leftrightarrow 3$ are located in a
cavity (Fig. 1) tuned to the resonance with transition $1
\leftrightarrow 2$.

\begin{figure}%[!ht]
  \includegraphics[scale=0.5,angle=0]{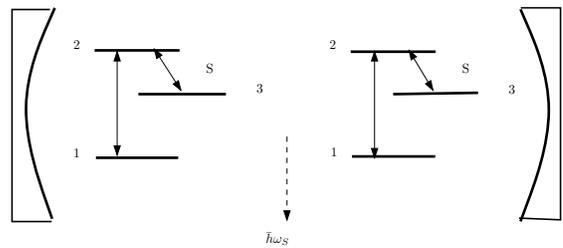}\\
  \caption{Scheme of transitions in two three-level $\Lambda$-type
  atoms in a cavity. Transition $1 \leftrightarrow 2$ is resonant with the cavity
  (Pumping) field, while S ($2 \leftrightarrow 3$) corresponds to the transition with creation
  of Stokes photon. The dashed arrow shows the discarding of the Stokes photon.}\label{f:1}
\end{figure}

If initially both atoms are in the ground state and cavity
contains one photon, than absorption of the photon by either atom
leads to creation of CE atomic state
\begin{eqnarray}
|\psi^{(12)}_{CE} \rangle = \frac{1}{\sqrt{2}} (|2 \rangle_I
\otimes |1 \rangle_{II} +|1 \rangle_I \otimes |2 \rangle_{II} ),
\label{13}
\end{eqnarray}
where $|n \rangle_j$ denotes the state of $j$-th atom. This state
manifests maximal amount of quantum fluctuations of the local
basic observables (Pauli operators)
\begin{eqnarray}
\sigma_x^{(j)}= |2 \rangle_j \langle 1|+H.c., \quad
\sigma_y^{(j)}=-i|2 \rangle_j \langle 1|+H.c., \nonumber \\
\sigma_z^{(j)}=|2 \rangle_j \langle 2|-|1 \rangle_j \langle 1|,
\nonumber
\end{eqnarray}
so that
\begin{eqnarray}
\mathbb{V}(\psi^{(12)}_{CE})=6. \nonumber
\end{eqnarray}
The corresponding energy of the system is
\begin{eqnarray}
E^{(12)}=\epsilon_2 \sim \hbar \omega_C , \label{14}
\end{eqnarray}
where $\epsilon_j$ denotes the energy of the corresponding atomic
level with respect to the ground state ($\epsilon_1=0$) and
$\omega_C$ is the cavity mode frequency.

This state (13) is unstable. There are the two channels of decay
of the excited atomic state:
\begin{eqnarray}
|2 \rangle_j \rightarrow \left\{ \begin{array}{ll} |1 \rangle_j &
\mbox{with creation of cavity photon} \\ |3 \rangle_j & \mbox{with
creation of Stokes photon} \end{array} \right. \nonumber
\end{eqnarray}
The first way returns the system into the initial state. After
that, the process would be repeated. The second decay channel
creates the new CE state
\begin{eqnarray}
|\psi^{(13)} \rangle = \frac{1}{\sqrt{2}} (|3 \rangle_I \otimes |1
\rangle_{II}+|1 \rangle_I \otimes |3 \rangle_{II} ), \label{15}
\end{eqnarray}
which manifests the same amount of quantum fluctuations as (12)
but with respect to the new local basic observables
\begin{eqnarray}
\sigma_x^{(j)}= |3 \rangle_j \langle 1|+H.c., \quad
\sigma_y^{(j)}=-i|3 \rangle_j \langle 1|+H.c., \nonumber \\
\sigma_z^{(j)}=|3 \rangle_j \langle 3|-|1 \rangle_j \langle 1|.
\nonumber
\end{eqnarray}
The corresponding energy is
\begin{eqnarray}
E^{(13)}= \epsilon_3+ \hbar \omega_S \sim \epsilon_2 , \nonumber
\end{eqnarray}
where $\omega_S$ denotes the frequency of Stokes photon. This is
the same energy as for the $(12)$ configuration (13).  If the
Stokes photon is now discarded, the energy is decreased
\begin{eqnarray}
E^{(13)} \rightarrow E^{(13)}_{min}= \epsilon_3 \nonumber
\end{eqnarray}
and the state (15) becomes stable (at least, with respect to the
dipole transitions). To discard the Stokes photon, we can think
either about its absorption by the cavity walls or about its free
leakage out of the cavity. In the latter case, detection of the
Stokes photon outside the cavity signalizes the creation of the
robust atomic entangled state (15). For further discussion of the
above scheme, see Biswas and Agarwal 2004, \c{C}ak{\i}r {\it et
al} 2004, \c{C}ak{\i}r {\it et al} 2005.

\section{Conclusion}

Summarizing, we should stress the generality of definition of CE
states has been discussed in Sec. 2. Physically it associates CE
with special behavior of expectation values of basic observables
and, in that way, with the maximal amount of quantum fluctuations.
In a sense, it follows Bell's ideology ( Bell 1966) that
entanglement manifests itself in local measurements and their
correlations. The possible role of quantum fluctuations in
formation of entangled states was also noticed by G\"{u}hne {\it
et al} (G\"{u}hne {\it et al} 2002) and Hofmann and Takeuchi
(Hofmann and Takeuchi 2003).

Since the classical level of description of physical systems
neglects existence of quantum fluctuations, the total variance (1)
can be chosen as a certain measure of {\it remoteness} of quantum
reality from classical picture. Thus, the coherent states with
minimal amount of quantum fluctuations are the closest states to
classical picture, while CE states represent the most nonclassical
states. In particular, Klyachko (Klyachko 2002) and Barnum {\it et
al} (Barnum {\it et al} 2003) have associated generalized coherent
states with the separable states of multipartite systems.

From the physical point of view, the definition, connecting CE
with quantum fluctuations reveals the way of preparing robust
entanglement (Sec. 4).

The general approach to quantum entanglement has been discussed in
Sec. 2 is based on the consideration of the symmetry properties of
physical systems. In particular, it associates definition of CE
with the orthogonal basis of the Lie algebra, corresponding to the
Lie group of the dynamical symmetry of the system. As it has been
shown in Sec. 3, this causes a certain relativity of quantum
entanglement with respect to the choice of the dynamic symmetry.
As an example, a single-qutrit entanglement was considered.

In the case of a two-qutrit system, the entanglement takes place
both in the $\mathrm{SU}(3)$ and $\mathrm{SU}(2)$ sectors. Since
the set of the $\frak{su}(3)$ basic observables (12) contains the
spin-1 operators, CE of two qutrits in the $\mathrm{SU}(3)$ sector
involves CE in the $\mathrm{SU}(2)$ sector but not vice versa. The
CE states of two spin-1 objects can be examined through the use of
the following symmetry relation
\begin{eqnarray}
\mathrm{SU}(2) \times \mathrm{SU}(2) \simeq SO(4). \nonumber
\end{eqnarray}

The symmetry based approach to quantum entanglement leads to a
certain ``stratification" of possible states of quantum systems
(Klyachko 2002). Namely, if $G$ is the dynamic symmetry group, the
SLOCC are defined by the  action of complexified group $G^c$.
Then, the different classes of states are given by the {\it
orbits} of the action of $g^c \in G^C$ in the Hilbert space
$\mathcal{H}_S$.

For example, in the case of three qubits, the dynamic symmetry of
the system is described by the Lie group
\begin{eqnarray}
\mathrm{SU}(2) \times \mathrm{SU}(2) \times \mathrm{SU}(2).
\nonumber
\end{eqnarray}
Thus, SLOCC belong to the group $SL(2,\mathbb{C})$. The orthogonal
basis of the corresponding Lie algebra $s \ell (2, \mathbb{C})$ is
given by the Pauli operators. It was shown by Miyake (Miyake 2003)
through the use of the mathematical analysis of multidimensional
matrices and determinants by Gelfand {\it et al} (Gelfand {\it et
al} 1994) that there are only four SLOCC nonequivalent classes of
states shown in the Table below.

\vspace{3mm}
\begin{center} Table 1.

\vspace{3mm}
\begin{tabular}{|l|l|}
  \hline
  $\frac{1}{\sqrt{2}} (|000 \rangle +|111 \rangle)$ & GHZ state \\
  \hline
  $\frac{1}{\sqrt{3}} (|001 \rangle +|010 \rangle +|100 \rangle)$ & W-state
  \\ \hline
  $\frac{1}{\sqrt{2}} \times \left\{ \begin{array}{l} (|001 \rangle +|010 \rangle)
  \\
  (|001 \rangle +|100 \rangle) \\ (|010 \rangle +|100 \rangle) \end{array} \right.$ & biseparable states
  \\ \hline
  $|000 \rangle$ & completely separable states \\
  \hline
\end{tabular}
\end{center}

\vspace{3mm}Similar classification was proposed by A\'{c}in {\it
et al} (A\'{c}in {\it et al} 2001) through the use of tripartite
witnesses.

Besides the classification, the notion of complex orbits allows to
introduce a proper measure $\mu$ of entanglement as the length of
minimal vector in the complex orbit (Klyachko 2002, Klyachko and
Shumovsky 2004). Note that all natural measures of entanglement
should be represented by the {\it entanglement monotones}, i.e. by
functions decreasing under SLOCC (Vidal 2000, Eisert {\it et al}
2003, Verstraete {\it et al} 2003) and that the above measure
obeys this condition.

In the case of an arbitrary pure two-qubit state
\begin{eqnarray}
|\psi_{2,2} \rangle = \sum_{\ell , \ell' =0}^1 \psi_{\ell \ell'}
|\ell \rangle \otimes |\ell' \rangle, \quad \sum_{\ell , \ell'}
|\psi_{\ell \ell'}|^2=1, \nonumber
\end{eqnarray}
the measure of entanglement $\mathcal{C} = \det [\psi]$, where
$[\psi]$ is the $(2 \times 2)$ matrix of the coefficients of the
state $|\psi_{2,2}\rangle$. This determinant represents the only
entanglement monotone in this case. To within a factor, this
measure coincides with the {\it concurrence} $\mathcal{C}(\psi)$
(Wootters 1998), which is usually used to quantify entanglement in
two-qubit systems:
\begin{eqnarray}
\mathcal{C}(\psi)=2|\det [\psi]|. \nonumber
\end{eqnarray}

In the case of three qubits, the measure is given by the absolute
value of Cayley hyperdeterminant multiplied by four (Miyake 2003)
also known as {\it 3-tangle} (Coffman {\it et al} 2000)
\begin{eqnarray}
\tau = 4|\psi_{000}^2 \psi_{111}^2+ \psi_{001}^2 \psi_{110}^2 +
\psi_{010}^2 \psi_{101}^2 + \psi_{100}^2 \psi_{011}^2 \nonumber \\
-2(\psi_{000} \psi_{001} \psi_{110} \psi_{111}  +\psi_{000}
\psi_{010} \psi_{101} \psi_{111} \nonumber \\ + \psi_{000}
\psi_{100} \psi_{011} \psi_{111} + \psi_{001} \psi_{010}
\psi_{101} \psi_{110} \nonumber \\ + \psi_{001} \psi_{100}
\psi_{011} \psi_{110} + \psi_{010} \psi_{100} \psi_{011}
\psi_{101} ) \nonumber \\ +4( \psi_{000} \psi_{011} \psi_{101}
\psi_{110}  + \psi_{001} \psi_{010} \psi_{100} \psi_{111})|,
\label{16}
\end{eqnarray}
where $\psi_{i,j,k}$ are the coefficients of the normalized state
\begin{eqnarray}
|\psi_{2,3} \rangle = \sum_{i,j,k=0}^1 \psi_{ijk} |i \rangle
\otimes |j \rangle \otimes |k \rangle . \label{17}
\end{eqnarray}
This is the again the only entangled monotone for the states (17).
In the case of GHZ (Greenberger-Horne-Zeilinger) state (the first
row in Table 1), 3-tangle (16) has the maximal value $\tau
(GHZ)=1$. For all other states in the Table 1, it has zero value,
so that these states are unentangled.

This fact allows us to separate essential from the accidental in
the definition of quantum entanglement. For example, violations of
Bell's inequalities is often considered as a definition of
entanglement. The so-called W-states (the second row in Table 1)
violate Bell's inequalities (Cabello 2002). But as we have seen,
these states do not manifest entanglement (at least in the
tripartite sector).

In fact, violation of Bell's inequalities means the absence of
hidden variables (Bell 1966) and can be observed even in the case
of generalized coherent states (Klyachko 2002), which are
unentangled by definition.

Other definitions based on the nonsepsrability and nonlocality of
states also have a limited application. For sure, they are
meaningless in the case of a single spin-1 particle entanglement
have been considered in Sec. 3.

In this paper, we have considered entanglement of pure states. The
generalization of the approach on the case of mixed states meets
certain complications. The point is that the density matrix
contains classical fluctuations caused by the statistical nature
of the state together with quantum fluctuations. Their separation
represents a hard problem of extremely high importance. One of the
possible approaches consists in the use of the methods of
thermo-field dynamics (Takahashi Y and Umezawa H 1996), which
allows to represent a mixed state in terms of a pure state of
doubled dimension.

\section*{Acknowledgement}

The authors are grateful to A. Beige, M. Christandl, A. Ekert, Z.
Hradil, and V. Manko for useful discussions.

\end{document}